# Two-dimensional topological insulator state in cadmium arsenide thin films

Alexander C. Lygo, Binghao Guo, Arman Rashidi, Victor Huang, Pablo Cuadros-Romero and Susanne Stemmer[a]

Materials Department, University of California, Santa Barbara, California 93106-5050, USA

[a] Corresponding author. Email: stemmer@mrl.ucsb.edu

**Abstract**

Two-dimensional topological insulators (2D TIs) are a highly desired quantum phase but few materials have demonstrated clear signatures of a 2D TI state. It has been predicted that 2D TIs can be created from thin films of three-dimensional TIs by reducing the film thickness until the surface states hybridize. Here, we employ this technique to report the first observation of a 2D TI state in epitaxial thin films of cadmium arsenide, a prototype Dirac semimetal in bulk form. Using magnetotransport measurements with electrostatic gating, we observe a Landau level spectrum and quantum Hall effect that are in excellent agreement with those of an ideal 2D TI. Specifically, we observe a crossing of the zeroth Landau levels at a critical magnetic field. We show that the film thickness can be used to tune the critical magnetic field. Moreover, a larger change in film thickness causes a transition from a 2D TI to a 2D trivial insulator, just as predicted by theory. The high degree of tunability available in epitaxial cadmium arsenide heterostructures can thus be used to fine-tune the 2D TI, which is essential for future topological devices.

Two-dimensional topological insulators (2D TIs) are a highly sought-after phase owing to their spin-polarized counterpropagating (helical) edge states, which are of great interest for their ability to host novel physical phenomena, such as the quantum spin Hall effect [1-3], and for topologically protected quantum computing [4]. 2D TIs are defined as possessing an inverted (relative to a normal semiconductor band structure) and gapped band structure which together, by the bulk-boundary correspondence, give rise to $\mathbb{Z}_2$ topological order and novel edge states [1]. Despite intensive research and a number of theoretically proposed systems [5], experimentally confirmed 2D TIs are, however, very rare. Materials for which experimental signatures indicative of a 2D TI state have been reported include quantum well structures [6-8] and monolayer van der Waals compounds [9, 10].

An alternative, and largely unexplored, route to a 2D TI is thin films of three-dimensional (3D) TIs whose film thickness is reduced such that there is a spatial overlap of the surface state wavefunctions [11-13]. In this case, the 3D TI's surface states gap out, forming degenerate *massive* Dirac states, and the material is a 2D TI if there is inversion between the confinement induced electron-like and hole-like subbands [13, 14]. An additional requirement is the absence of a strong potential difference between the top and bottom surfaces of the film (so called structural inversion asymmetry or SIA) as this introduces additional coupling between the massive Dirac states that, if strong enough, destroys the 2D TI phase [13, 15, 16]. Furthermore, as a function of film thickness, it is predicted that changes in the subband ordering can cause transitions between a 2D TI and 2D trivial insulator states. The thicknesses when these transitions are expected to occur depend sensitively on band parameters and microscopic details of the system [12]. Hence, it is of interest to explore high quality thin films of 3D TIs prepared with controlled thicknesses and negligible SIA.

In this study, we employ the confinement approach to epitaxial thin films of cadmium arsenide ($Cd_3As_2$). While $Cd_3As_2$ is a prototype 3D Dirac semimetal with large band inversion in bulk [14, 17-20], it is a nearly ideal 3D TI in thin films [21, 22] with high surface state mobility and no parasitic bulk conduction at low temperature. As the $Cd_3As_2$ film thickness is further reduced, theory predicts a transition to a 2D TI (quantum spin Hall insulator) with a wide energy gap [14]. Recently, we reported evidence of surface state hybridization in a 20 nm-thin $Cd_3As_2$ thin film [23], an essential step towards a 2D TI. Here, we present the first experimental evidence of a 2D TI state in-(001)-oriented $Cd_3As_2$ films. To this end, we use Landau level spectroscopy, which, as we will discuss next, can unambiguously identify surface state hybridization and the inversion of the bands from the behavior of the zeroth Landau levels in high-quality films, as well as other essential details, such as SIA. Moreover, just as predicted by the theoretical models, we show that a small (6 nm) change in film thickness causes a transition from a 2D TI to a 2D trivial insulator.

A hallmark of a 2D TI in a perpendicular magnetic field ($B$) is a *crossing* of two zero-energy ($n = 0$) Landau levels at a critical field ($B_c$), as shown in Fig. 1(a), and a zero conductance ($v = 0$) quantum Hall plateau when the chemical potential lies in the energy gap between the $n = 0$ Landau levels (after refs. [15, 16]; for details of the calculations, see the Supplementary Materials [24]). Because of band inversion, one of the two $n = 0$ Landau levels is electron-like and originates from the valence band and the other is hole-like and originates from the conduction band. At $B_c$, the system undergoes a phase transition from a nontrivial insulator to a trivial one. Concurrently, the $v = 0$ quantum Hall plateau disappears but then reemerges for $B > B_c$. The zero energy Landau level crossing and re-entrant quantum Hall effect at $B_c$ are robust signatures of a 2D TI state [6], while its other key feature, the helical edge states and quantum spin Hall effect, are easily obscured by trivial edge conduction paths [27, 28] and by their extreme sensitivity to disorder [29-31]. The zeroth Landau level crossing at $B_c$ distinguishes the 2D TI from all other potential states of a topological thin film. For example, Fig. 1 shows calculated [15, 16] Landau fan diagrams of several other possible states [24], such as the Landau level spectrum of a hybridized 3D TI *without* band inversion (a film in the trivial thickness regime, for example) [Fig. 1(b)]. Here, because band inversion is absent, all electron-like Landau levels originate from the conduction band and all hole-like ones from the valence band. Thus, the $n = 0$ Landau levels never cross and the $v = 0$ quantum Hall persists and widens with increasing $B$. If the film is in the nontrivial thickness regime but strong SIA is present, the crossing of the $n = 0$ Landau levels of the 2D TI becomes an anti-crossing [Fig. 1(c)], and the system is also a trivial insulator. Additionally, the $v = 0$ quantum Hall plateau is present at all $B$ but, in contrast to the preceding case, its width evolves nonmonotonically with increasing $B$. Finally, Fig. 1(d) shows that a $v = 0$ quantum Hall state can also be observed for a *non-hybridized* (thick) 3D TI film when there is SIA and when the chemical potential lies between the $n = 0$ Landau levels of the top and bottom surfaces [32]. In this case, the $n = 0$ Landau levels are non-dispersing, as has indeed been observed for several 3D TIs [33-36], and the width of the $v = 0$ quantum Hall plateau is constant in $B$.

The presence of SIA can also be detected by characteristic crossings of Landau levels at higher energies, as seen in Figs. 1(c) and 1(d). This latter feature results in complicated filling factor ($v$) sequences in the quantum Hall effect as a function of carrier density and magnetic fields, as observed for thicker (~ 50 nm) $Cd_3As_2$ films that are in the 3D TI state [21]. No such crossings occur without SIA [see Figs. 1(a) and 1(b)].

Combined, these distinguishing features, especially the dispersion of the $n = 0$ Landau levels and a reentrant $v = 0$ quantum Hall plateau, provide experimental signatures of the four different possible phases in thin films. Clearly, it is essential to tune the Fermi level to charge neutrality and into the gap between the $n = 0$ Landau levels to distinguish these phases.

To study the electronic state of very thin $Cd_3As_2$ films, we performed low temperature magnetotransport measurements on top gated Hall bar structures. Details about sample growth, device fabrication, transport measurements can be found in the Supplementary Information [24]. Figures 2 (a,b) show the longitudinal ($\sigma_{xx}$) and Hall conductivity ($\sigma_{xy}$), respectively, of a 20 nm (001)-oriented $Cd_3As_2$ film (sample 1), calculated by tensor inversion from the resistivities (shown in the Supplementary Information [24]), as a function of applied top gate bias, $V_g$, and $B$ (for a plot of $V_g$ versus carrier density, see Supplementary Information [24]). The labels in Fig. 2(a), marking the $\sigma_{xx}$ minima, denote the filling factors of the corresponding integer quantum Hall plateaus in Fig. 2(b). We observe a sequence of Landau levels that produce well defined quantum Hall plateaus with both even and odd $\nu$. Taken alone, these filling sequences would be consistent with any of the states in Fig. 1. As discussed above, the low energy portion of the spectrum, around $\nu = 0$, is crucial to distinguish them.

We thus first turn our attention to the region around charge neutrality ($-1.5 > V_g > -2$ V). The first notable feature is the existence of a gap at zero $B$, as evidenced by the very low conductivity ($\sigma_{xx} \cong 0.06\ e^2/h$) and insulating behavior (the temperature dependence is discussed below). Moreover, the salient feature here are two Landau levels that originate at different $V_g$ and, as $B$ is increased, converge, meeting at approximately $B_c = 9.4$ T, before diverging for larger $B$. Concurrently, the $\nu = 0$ quantum Hall plateau vanishes and then remerges. This latter feature can be seen especially clearly in the $\sigma_{xy}$ traces shown in Fig. 2(b): here, $\nu = 0$ is present at low $B$ (yellow-green traces), absent at intermediate $B$, and reentrant at $B > 10$ T (dark blue traces) (for additional clarity, see also the Supplementary Material for a zoom in of the $\sigma_{xy}$ traces around the $\nu$

= 0 plateau [24]). Accordingly, we identify the Landau levels crossing at $B_c$ = 9.4 T as the $n$ = 0 levels. In the regions where these two Landau levels are well separated, $\sigma_{xx}$ is approximately zero and $\sigma_{xy}$ plateaus at zero. By contrast, where $\sigma_{xx}$ is finite (around $B_c$), $\sigma_{xy}$ changes sign smoothly (Fig. S5 [24]). *The Landau level fan diagram is thus in perfect agreement with that of an idealized prototype 2D TI shown in Fig. 1(a)*. It is *not* in agreement with either of the possible other states in Fig. 1. We note that it is the significant dispersion (in $B$) of both $n$ = 0 levels, and the absence of any other states nearby, that allows for the clear re-entrant $v$ = 0 quantum Hall plateau. By comparison, for HgTe quantum wells the *p*-type $n$ = 0 quantum Hall plateau is nearly non-dispersive (indicative of a low Fermi velocity) and has a more complex dispersion [37, 38], causing the re-entrant quantum Hall effect to be rather complex (e.g., $v = 0 \rightarrow v = 1$ (-1) $\rightarrow v = 0$ [6]).

Next, we discuss the Landau level spectrum away from the charge neutrality point, important for understanding the stability of the 2D TI state in this film. There are two Landau levels, which are marked with arrows in Fig. 2(a) from a higher energy, bulk-originated subband. This is evident by the fact that they can be traced back to a different point on the $V_g$ axis (see Fig. S6 where lines have been drawn for clarity to show the intercepts of the Landau levels with the $V_g$ axis [24]). When the chemical potential crosses either of these additional Landau levels, $v$ changes by 1 [see corresponding region in Fig. 2(b)], indicating that they are non-degenerate. The existence of electronic subbands is consistent with expectations for a quantum confined thin film [12, 14].

The second important feature of the Landau level spectrum is the fact, that, with the exception of the $n$ = 0 Landau levels, Landau levels that originate from the electronic states near charge neutrality do *not* cross. As discussed above, the *absence* of any other crossings in the Landau fans to which these $n$ = 0 levels belong shows that there is no perceptible energy offset (SIA), although it is present for thicker films in similar structures [21, 22]. While it is perhaps not surprising that a 20 nm thin film with a small gap does not sustain a potential offset, the absence of strong SIA is an essential pre-requisite for the observed 2D TI state. Finally, in the region $V_g <$ -2 V we observe a deep minimum in $\sigma_{xx}$ and a plateau in $\sigma_{xy}$ corresponding to $\nu = -1$. At more negative $V_g$ we do not observe additional quantum Hall plateaus, possibly because the chemical potential resides in an electronic band of the buffer layer, causing spillover of carriers, or because of a high density of Landau levels from the more heavy, lower energy, valence band states (e.g., similar to HgTe quantum wells).

To demonstrate the sensitivity of the 2D TI state to small changes in film thickness, we performed additional transport measurements on 18 nm, 19 nm, and 22 nm $Cd_3As$ films prepared under nominally identical conditions as sample 1 (see Fig. 3). The main result is a qualitatively identical Landau level spectrum. The primary quantitative differences as a function of film thickness are: (i) with increasing film thickness $B_c$ increases, which is similar to the thickness dependence observed in HgTe quantum wells [6], and (ii) the bulk-originated subband moves to higher $V_g$, consist with expected behavior for quantum confined thin films. The similarity of the four Landau level spectra is in agreement with our previous observations, namely a qualitative consistency of *higher energy Landau levels* ($n > 0$) behavior with thickness [39]. However, we now see clearly the opening of a hybridization gap, which is most dramatically displayed in the behavior of the zero energy Landau levels, whose behavior extremely sensitive to the thickness.

As the film thickness is further decreased, it is predicted [14] that changes in the subband ordering cause a transition from 2D TI to a 2D trivial insulator. Figure 4 shows $\sigma_{xx}$ and $\sigma_{xy}$ versus $V_g$ and $B$ for a 14 nm film prepared under similar conditions as those discussed above. Qualitatively, away from the charge neutrality point [Figs. 4(a) and 4(b)], the data is strikingly similar to the other samples. Quantum Hall plateaus with both even and odd $\nu$, indicating a degeneracy lifting, are present, but no crossings of higher energy Landau levels in the low energy fans are evident, demonstrating again the absence of SIA. Distinct from the other films, the Landau levels that originate from states of a higher energy subband are absent in the $V_g$ range studied here, consistent with the expectation that the subband ordering changes as a function of film thickness. The important difference of this film is seen near charge neutrality. For all samples, there is a clear insulating state at $B = 0$ accompanied by $\sigma_{xy} = 0$, indicating hybridization of the surface states; for the 14 nm film, however, the $n = 0$ Landau levels *diverge* and the $v = 0$ quantum Hall plateau widens with increasing magnetic field. The Landau level spectrum of the 14 nm film is in remarkably good agreement with that of a 2D trivial insulator shown in Fig. 1(b). We conclude from this that a 2D TI state in $Cd_3As_2$ can be achieved in 18-22 nm films while a small reduction in the film thickness (to 14 nm) causes a transition from a 2D topological insulator to a trivial one, consistent with the change in the subband ordering that is evident in the higher energy spectrum. Small discrepancies between the thickness ranges for the different phases in the observations vs. predictions (ref. [14]) can easily be explained by the fact the microscopic details of the heterostructures, which determine important parameters such as the Fermi velocity, were not considered in the models.

In summary, we have observed the hallmarks of a nearly ideal 2D TI state in thin films of (001)-oriented $Cd_3As_2$, including insulating behavior at zero field and an energy gap between two $n = 0$ Landau levels that closes at $B_c$. The zeroth Landau level crossing is remarkably well resolved, aided by relatively high Fermi velocities of the electronic states that give rise to the zero-energy Landau levels, low disorder, and a good separation in energy from, e.g., a high density of low mobility valence band states found in other systems [37, 40]. These results establish that thin films of $Cd_3As_2$ constitute a new member of the very small, highly sought-after family of 2D TIs discovered to date. Crucially, this 2D TI state is realized via a previously theoretically suggested route, namely by quantum confinement. We also demonstrated that reducing the film thickness further induces a transition to a 2D trivial insulator, also consistent with theoretical predictions. The wide range of additional heterostructure parameters that can be tuned, such as film strain, makes the 2D TI phase in $Cd_3As_2$ films extraordinarily tunable. This tunability could prove extremely useful in designing and testing future superconducting hybrid junctions for quantum information systems, which depend on finely tuned energy scales, and novel correlated states [41]. This study provides clear directions for the future work such as more detailed study of the thickness dependence of the electronic state of $Cd_3As_2$ films, particularly one that includes ultrathin (< 10 nm) films. Finally, the results presented here demonstrate the possibility of realizing the quantum spin Hall effect in thin films of $Cd_3As_2$ and a next step should be investigations of the edge states physics, which requires smaller devices and low defect density mesa boundaries.

## Acknowledgements

The authors are grateful to Andrea Young and Xi Dai for very helpful discussions. The research was supported by the Air Force Office of Scientific Research (Grant No. FA9550-21-1-0180) and by the Office of Naval Research (Grant No. N00014-21-1-2474). A.C.L and B.G. also thank the Graduate Research Fellowship Program of the U.S. National Science Foundation for support (Grant Nos. 1650114 and 2139319). This work made use of the MRL Shared Experimental Facilities, which are supported by the MRSEC Program of the U.S. National Science Foundation under Award No. DMR 1720256.

**Appendix A: Temperature dependent conductance at ν = 0 in the 2D TI state**

To further investigate the 2D TI, including the gapped state at low B, we performed temperature dependent two-point conductance (G) measurements, shown in the Supplementary Information [24]. Figure S13 [24] shows the temperature dependence of the minimum G of film 1 between the $n = 0$ Landau levels at $B < 8$ T [Fig. S13(a)], around $B_c$ [Fig. S13(b)], and for $B > 12$ T [Fig. S13(c)]. For $B < 8$ T and $B > 12$ T, respectively, G shows an exponential-like increase with increasing temperature, consistent with a gapped state. In these two regimes, the observed temperature dependence can be described $G(T) = G_0 \exp[-(T_0/T)^p]$, where $G_0$ is a temperature independent prefactor, $T_0$ is the characteristic hopping temperature and $p$ is a model parameter that depends on the density of states at the Fermi level and takes the values $0 < p < 1$ (see Supplementary Information [24]). A value of $p = 1$ corresponds to Arrhenius behavior and $T_0 = E_a/k_B$ where $E_a$ is the activation energy and $k_B$ is Boltzmann's constant. We find that for $B < 8$ T [Fig. S13(a)], the temperature dependence of G can be described by $p = 1/3$, i.e., 2D Mott variable range hopping (VRH), over some temperature range (see dashed lines). At intermediate B (7 T $< B <$ 13 T), G shows dramatically different behavior [Fig. S13(b)]. Within this range, but away from $B_c$, G is a nonmonotonic function of temperature with $dG/dT > 0$ at low temperatures that transitions to T-linear dependence with a negative slope at higher temperatures. At $B_c$ ($B = 9$ T and $B = 10$ T traces), G shows metallic behavior (it increases monotonically with decreasing temperature) and is approximately linear in temperature above 12.5 K. For $B \geq 13$ T [Fig. S13(c)], the temperature dependence roughly follows Arrhenius behavior above 3.5 K, consistent with a clean, gapped state. Most importantly, the crossover from insulating to metallic temperature dependence occurs around $B = B_c$, consistent with a *crossing* of the $n = 0$ Landau levels. Near $B_c$, the maximum value of $\sigma_{xx}$ is ~ 0.85 $e^2/h$ (see Supplementary Information Fig. S14 for $\sigma_{xx}$ values between the $n = 0$ Landau levels [24]). Interestingly, this $\sigma_{xx}$ value is close to twice of the universal value of 0.5 $e^2/h$ [42], consistent with the crossing of *two* Landau levels – by contrast, at the other quantum Hall transitions, $\sigma_{xx}$ is close to 0.5 $e^2/h$. (we note that $G$, presented in Fig. S13, is smaller than $\sigma_{xx}$, shown, due to the contribution of contact/series resistances, including the ungated regions near the contacts). The origin of the "strange metal" (*T*-linear) behavior at the crossing of hole-like and electron-like zeroth Landau levels and, more generally, the nature of this "Dirac-like" state, warrants further investigations, including theoretically.

**Appendix B: Detectability of the helical edge states ion the 2D TI state**

We briefly comment here on the detectability of helical edge states, expected to be present for $B < B_c$. It is generally accepted in the 2D TI literature [9, 43, 44] that much smaller devices than those studied here, with carefully prepared edges [31], are needed to characterize the very fragile helical edge states, because the dimensions of the devices studied here exceed the phase coherence length (~1 μm). In devices larger than this dimension, it is generally found that $\sigma_{xx}$ is smaller than $2e^2/h$ at all temperatures [9, 43, 44]. Future experiments will address fabrication challenges for smaller devices. Already in these large devices, however, the data hints at potentially rich physics. For one, the deepest minimum of $\sigma_{xx}$ occurs at $B = 2.4$ T [$\sigma_{xx}(2.4\ \text{T}) = 0.008\ e^2/h$] and not at $B = 0$ [$\sigma_{xx}(0\ \text{T}) = 0.06\ e^2/h$], where the energy gap is largest (see Supplementary Information Fig. S14 [24]). This is also reflected in the temperature dependence of $G$, because $T_0$ is largest (125±8 K) at $B = 2$ T. One possible interpretation is that the decrease in $\sigma_{xx}$ away from $B = 0$ reflects increased scattering of the edge states due to time-reversal symmetry breaking [45]. Competition between this enhanced scattering and the closing gap causes the minimum of $\sigma_{xx}$ to occur at finite $B$. Secondly, VRH hopping does not completely describe the

data – at the lowest temperatures, we note that $G$ saturates (see Fig. S13a), possibly indicative of a second transport path.

**Figure Captions**

**Figure 1:** Characteristic Landau level spectra of different topological phases in thin films. (a) 2D TI, (b) 2D trivial insulator, (c) hybridized 3D TI with SIA, and (d) 3D TI with SIA. The labels denote quantum Hall filling factors. In (a), (b) and (c) electron-like (hole-like) levels are show in orange (blue) and in (d) Landau levels from the higher energy surface are shown in purple and those from the other surface are shown in green. In all, the $n = 0$ Landau levels are drawn with heavier line weight. All four phases can produce a $v = 0$ quantum Hall state (a 3D TI without SIA has degenerate $n = 0$ Landau levels and there is no $v = 0$ quantum Hall state). See Supplementary Materials for details of the calculations.

**Figure 2:** Landau levels and quantum Hall effect of sample 1 (20 nm film). Magnetic field ($B$) and gate voltage ($V_g$) dependence of the (a) longitudinal ($\sigma_{xx}$) and (b) Hall ($\sigma_{xy}$) conductivities of sample 1. The labels in (a) denote the corresponding quantum Hall filling factors in (b). The black arrows mark additional Landau levels from a higher energy subband.

**Figure 3:** Magnetic field ($B$) and gate voltage ($V_g$) dependence of the longitudinal ($\sigma_{xx}$) conductivity of (a) 18 nm, (b) 19 nm, and (c) 22 nm films.

**Figure 4:** Landau levels and quantum Hall effect of a 14 nm film. Shown are the $B$ and $V_g$ dependence of $\sigma_{xx}$ (a) and $\sigma_{xy}$ (b) of sample. Labels in (a) denote quantum Hall filling factors obtained from (b).

**Figures with captions**

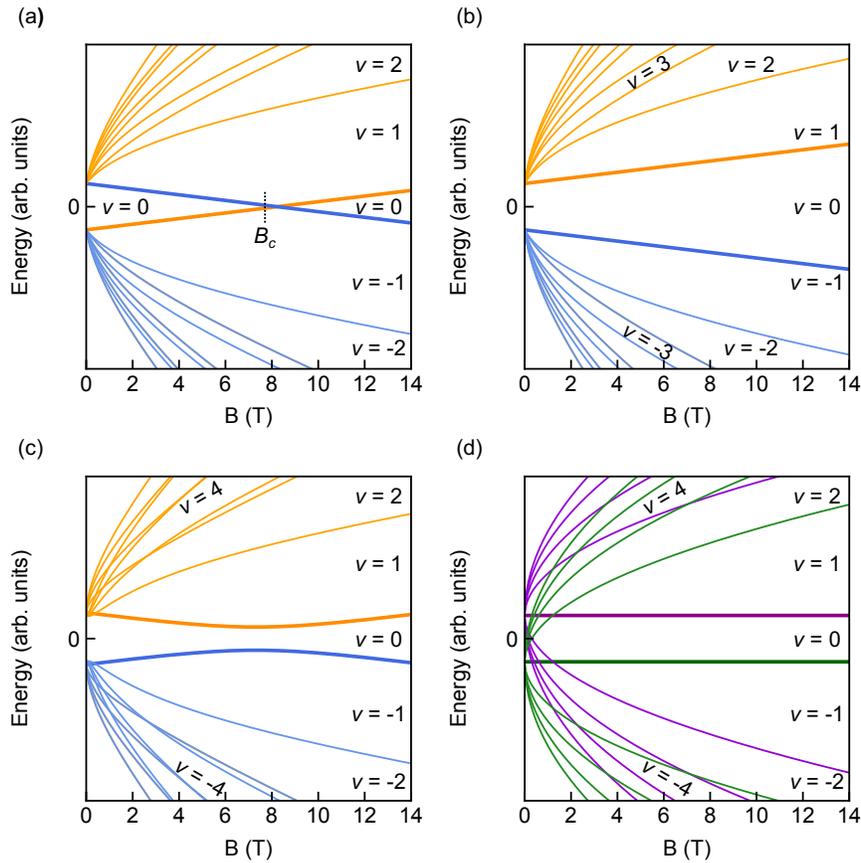

**Figure 1:** Characteristic Landau level spectra of different topological phases in thin films. (a) 2D TI, (b) 2D trivial insulator, (c) hybridized 3D TI with SIA, and (d) 3D TI with SIA. The labels denote quantum Hall filling factors. In (a), (b) and (c) electron-like (hole-like) levels are show in orange (blue) and in (d) Landau levels from the higher energy surface are shown in purple and those from the other surface are shown in green. In all, the $n = 0$ Landau levels are drawn with heavier line weight. All four phases can produce a $v = 0$ quantum Hall state (a 3D TI without SIA has degenerate $n = 0$ Landau levels and there is no $v = 0$ quantum Hall state). See Supplementary Information for details of the calculations.

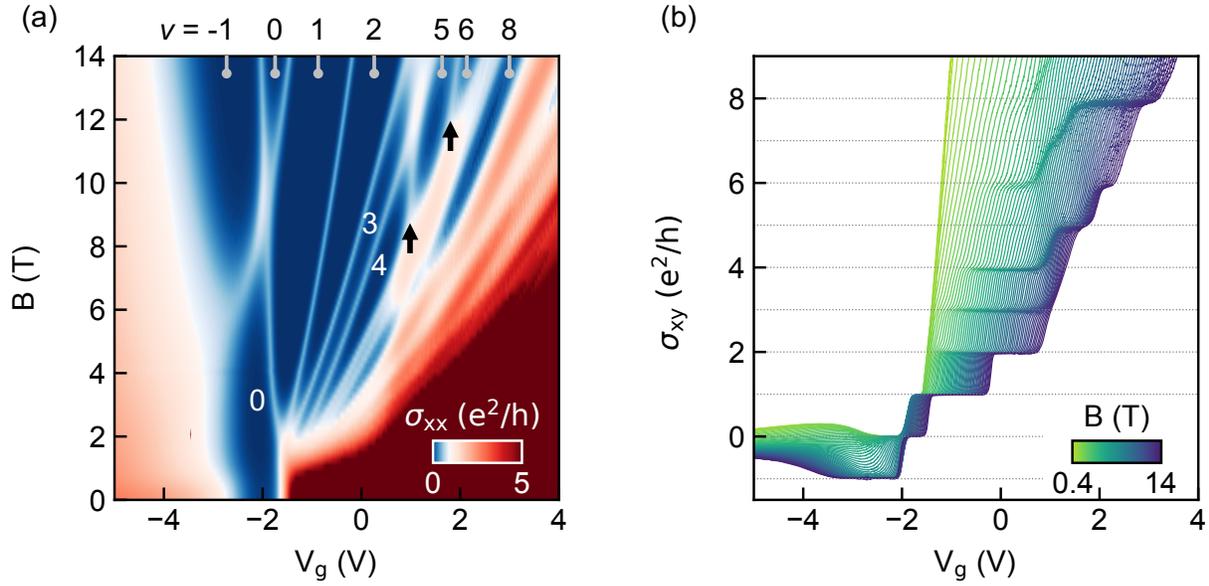

**Figure 2:** Landau levels and quantum Hall effect of sample 1 (20 nm film). Magnetic field ($B$) and gate voltage ($V_g$) dependence of the (a) longitudinal ($\sigma_{xx}$) and (b) Hall ($\sigma_{xy}$) conductivities of sample 1. The labels in (a) denote the corresponding quantum Hall filling factors in (b). The black arrows mark additional Landau levels from a higher energy subband.

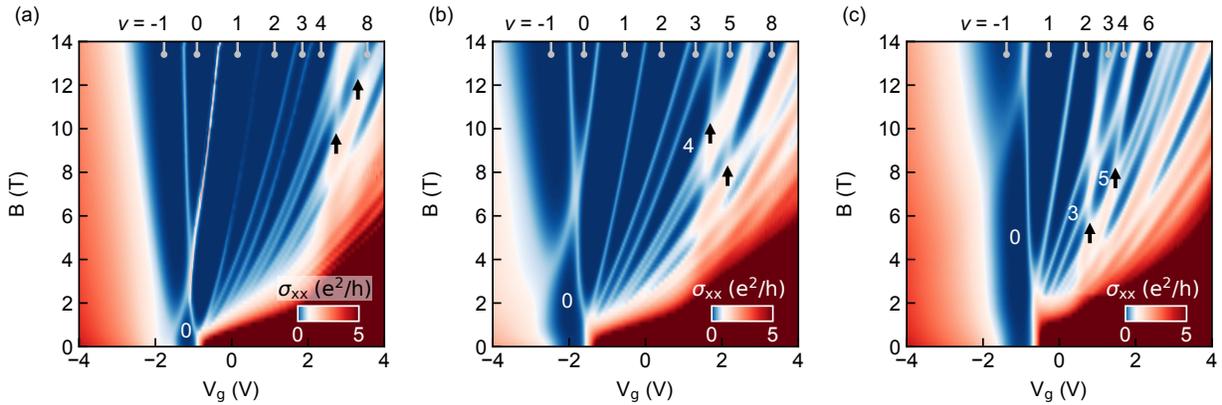

**Figure 3:** Magnetic field ($B$) and gate voltage ($V_g$) dependence of the longitudinal ($\sigma_{xx}$) conductivity of (a) 18 nm, (b) 19 nm, and (c) 22 nm films.

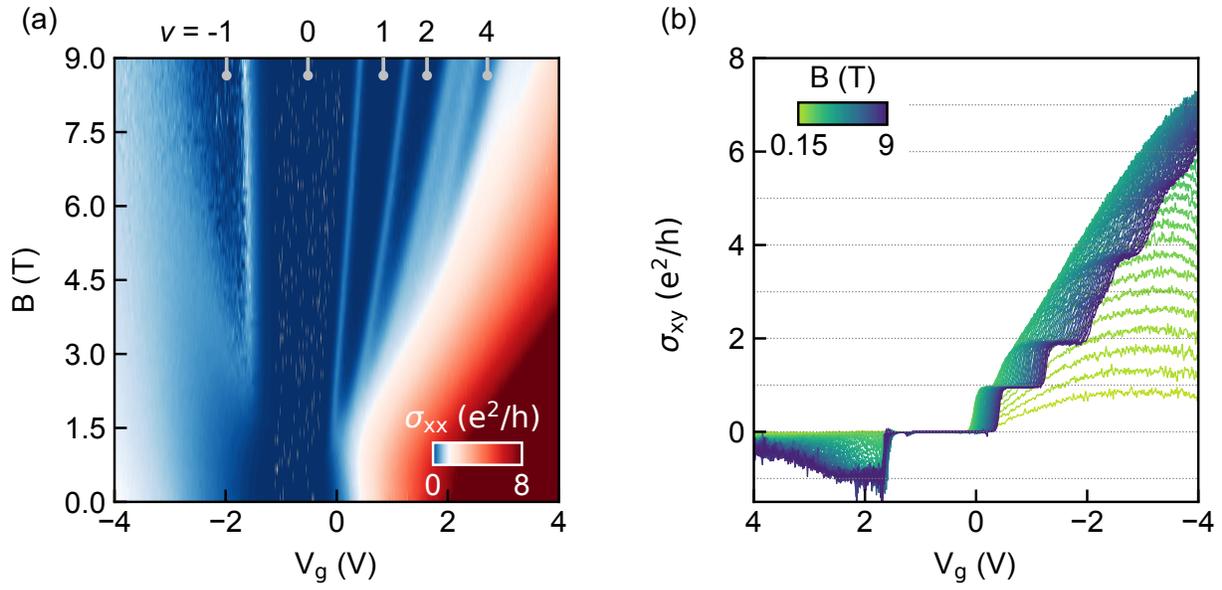

**Figure 4:** Landau levels and quantum Hall effect of a 14 nm film. Shown are the $B$ and $V_g$ dependence of $\sigma_{xx}$ (a) and $\sigma_{xy}$ (b) of sample. Labels in (a) denote quantum Hall filling factors obtained from (b).

# Supplementary Information

# Two-dimensional topological insulator state in cadmium arsenide thin films

**Alexander C. Lygo, Binghao Guo, Arman Rashidi, Victor Huang, Pablo Cuadros-Romero, and Susanne Stemmer**

Materials Department, University of California, Santa Barbara, California 93106-5050, USA

# Contents

# Model of a hybridized 3D TI

A two-dimensional topological insulator state can be realized in thin films of 3D topological insulators when the film thickness, $L$, is reduced such that there is a spatial overlap of the surface state wavefunctions. Starting from the bulk Hamiltonian of a 3D TI, the effective Hamiltonian in the thin film limit was derived in Ref. [1], the result being:

$$H_{eff} = \begin{pmatrix} h_+(k) & V_0 I_{2\times 2} \\ V_0 I_{2\times 2} & h_-(k) \end{pmatrix}. \tag{S1}$$

Here, $h_\pm(k) = E_0 - Dk^2 - \hbar v_F(k_x \sigma_y + k_y \sigma_x) \pm \left(\frac{\Delta}{2} - Mk^2\right)\sigma_z$, where $\sigma_i$ are Pauli matrices in the basis of spin-up and spin-down, $k^2 = k_x^2 + k_y^2$, $v_F$ is the Fermi velocity, and $E_0, D, \Delta$, and $M$ are material specific band parameters. In the presence of structural inversion asymmetry (SIA), $2V_0$ is the potential difference between the top and bottom surface of the thin film. If $V_0 = 0$ (no SIA), $H_{eff}$ is simply the Benevig-Hughes-Zhang Hamiltonian [2]. The condition $\Delta M > 0$ describes a 2D TI. The term $\left(\frac{\Delta}{2} - Mk^2\right)$ is a mass term due to hybridization and the parameters $\Delta$ and $M$ depend on $L$ and approach zero in the large $L$ limit, returning the system to a 3D TI.

Landau level energies of a hybridized 3D TI were derived in Ref. [3]. Here we briefly summarize their results. With the magnetic field applied normal to the plane of the thin film, the Landau level energies are obtained from $H_{eff}$ by first making the substitution, $\vec{k} \to -i\nabla + e\vec{A}/\hbar$, where $e$ is the elementary change and $\hbar$ the reduced Plank's constant. Under the Landau gauge $\vec{A} = By\hat{x}$. Using ladder operators $a = \frac{l_B}{\sqrt{2}}(k_x - \partial_y - y/l_B^2)$ and $a^\dagger = \frac{l_B}{\sqrt{2}}(k_x + \partial_y - y/l_B^2)$, where $l_B = \sqrt{\hbar/eB}$ is the magnetic length and defining $\omega = 2M/l_B^2$, the effective Hamiltonian in a magnetic field reads:

$$H_{eff}(B) = \begin{pmatrix} \frac{\Delta}{2} - \omega\left(a^\dagger a + \frac{1}{2}\right) & i\frac{\sqrt{2}\hbar v_F}{l_B} a & V_0 & 0 \\ -i\frac{\sqrt{2}\hbar v_F}{l_B} a^\dagger & -\frac{\Delta}{2} + \omega\left(a^\dagger a + \frac{1}{2}\right) & 0 & V_0 \\ V_0 & 0 & -\frac{\Delta}{2} + \omega\left(a^\dagger a + \frac{1}{2}\right) & i\frac{\sqrt{2}\hbar v_F}{l_B} a \\ 0 & V_0 & -i\frac{\sqrt{2}\hbar v_F}{l_B} a^\dagger & \frac{\Delta}{2} - \omega\left(a^\dagger a + \frac{1}{2}\right) \end{pmatrix}. \tag{S2}$$

The term $E_0 - Dk^2$ is dropped from $h_\pm(k)$ as it does not significantly influence the Landau level energies. Taking the eigenstates, $|n\rangle$, of the quantum harmonic oscillator as a basis, the trial solutions for $n = 0$ are:

$$\psi_0 = \begin{pmatrix} 0 \\ \phi_{2,0}\langle y|0\rangle \\ 0\rangle \\ \phi_{4,0}\langle y|0\rangle \end{pmatrix}, \tag{S3}$$

and for $n > 0$ they are:

$$\psi_n = \begin{pmatrix} \phi_{1,n}\langle y|n-1\rangle \\ \phi_{2,n}\langle y|n\rangle \\ \phi_{3,n}\langle y|n-1\rangle \\ \phi_{4,n}\langle y|n\rangle \end{pmatrix}, \tag{S4}$$

where $\phi_{i,n} = C_{i,n}/\sqrt{L_x} e^{ikx}$ with $C_{i,n}$ being constants and

$$\langle y|n\rangle = \frac{1}{\sqrt{n! 2^n l_B \sqrt{\pi}}} \exp\left(-\frac{(y-y_0)^2}{2l_B^2}\right) \mathcal{H}_n\left(\frac{(y-y_0)}{l_B}\right), \tag{S5}$$

where $\mathcal{H}_n$ are the Hermite polynomials and $y_0$ is the guiding center of the wave packet. Using Eqs. S2, S3, and S4, the energies of the Landau levels for $n = 0$ were determined to be:

$$E_0^\pm = \pm\sqrt{\left(-\frac{\Delta}{2} + \frac{\omega}{2}\right)^2 + V_0^2}, \tag{S6}$$

and for $n > 0$ the energies are:

$$E_{n,s}^\pm = \pm\sqrt{\left(\varepsilon_n + s\sqrt{\left(\frac{\omega}{2}\right)^2 + V_0^2 \left(\frac{\sqrt{2n}\hbar v_F}{l_B}\right)^2 + V_0^2 \left(\frac{n\omega - \frac{\Delta}{2}}{\varepsilon_n}\right)^2}\right)^2} \tag{S7}$$

where $\varepsilon_n = \sqrt{n\left(\frac{\sqrt{2}\hbar v_F}{l_B}\right)^2 + \left(\frac{\Delta}{2} - \omega\right)^2}$. For the Landau level spectra show in the Fig. 1 of the main text we have used $v_F = 5 \times 10^5$ m/s (a realistic value for Cd$_3$As$_2$ thin films [4]) throughout. For the spectrum of a 2D TI (Fig. 1a), $\Delta$=-20 meV, $M$=-800 meVnm$^2$, and $V_0 = 0$. For the spectrum of a 2D trivial insulator without SIA (Fig. 1b), $\Delta$=-20 meV, $M = 800$ meVnm$^2$, and $V_0 = 0$. For the spectrum of a hybridized 3D TI with SIA (Fig. 1c) we used $\Delta = -20$ meV, $M = -900$ meVnm$^2$ and $V_0 = 10$ meV. For the spectrum of a 3D TI with SIA (Fig. 1d) we used $\Delta = 0$, $M = 0$, and $V_0 = 10$ meV. These values are chosen for illustrative purposes only, as realistic model parameters for Cd$_3$As$_2$ thin films on a substrate are not known.

**Sample growth**

The Cd$_3$As$_2$ thin films discussed in the main text were grown by molecular beam epitaxy (MBE). Samples were grown on epi-ready undoped (001) GaSb substrate offcut 1.5° or 3° towards (111)B. The substrates were annealed for 12 hours at 150 °C in a high-vacuum load lock chamber. The substrate's native oxide was removed in an ultra-high vacuum (UHV) preparation chamber using atomic hydrogen etching for 1 hour at a substrate temperature of 500 °C, as measured by the heater thermocouple or by thermal desorption under Sb$_2$ flux in the UHV growth chamber. In the following, all temperatures were measured using an optical pyrometer, unless otherwise stated. The buffer layer growth consisted of: 100 nm of GaSb grown at approximately 490 °C to smooth the surface followed by 500 nm or 1 µm of Al$_{0.45}$In$_{0.55}$Sb grown between 390 °C and 410 °C. Afterwards, the substrate was cooled to a substrate heater thermocouple temperature between 120 °C 170 °C to grow Cd$_3$As$_2$. To protect the Cd$_3$As$_2$ film surface, 3 nm of GaSb was grown as a capping layer at the Cd$_3$As$_2$ growth temperature.

**Device fabrication**

Eight-arm Hall bars were patterned using standard photolithographic techniques. Mesas were isolated by argon ion beam dry etching. Sample 1 Hall bars were contacted by sputtered Au/Ti (200/50 nm) leads or electron-beam evaporated Au/Pt/Ti (200/20/20 nm) leads. For top gates, ~25 nm AlO$_x$, grown by atomic layer deposition (ALD) using a 120 °C process, served as the dielectric. The dielectric thickness was determined using ellipsometry performed on a Si spectator chip from the same processing run. Immediately prior to AlO$_x$ deposition, the entire chip was exposed to a brief low energy oxygen plasma. The gate electrodes were deposited by thermal evaporation of Au/Ni (200/50 nm). The gate electrode coved the top and sides of the channel. Fig. S1 shows an optical micrograph of Hall bar fabricated from the 20 nm sample. The labels denote the contact configuration used for 4-probe measurements.

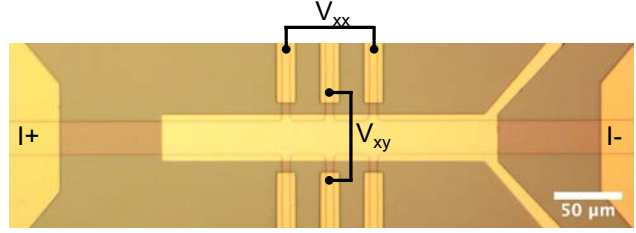

**Fig. S1.** Optical micrograph Hall bar fabricated from sample 1. The Cd$_3$As$_2$ appears reddish-pink. The labels denote the contact configuration used for 4-probe measurements.

**Transport measurements**

All transport measurements were performed in a Quantum Design PPMS with a base temperature $T = 1.8$ K, in magnetic fields $B$ up to 14 T. Temperature dependent measurements of the 20 nm were performed from 1.8 K to 40 K. Standard lock-in detection techniques were used for all measurements. Except for the 14 nm sample, 4-probe measurements were performed by applying a 10 mV AC (17.77 Hz) voltage to a ~10 M$\Omega$ resistor in series with the devices. The resulting current (0.15-1 nA) was measured using a Stanford Research Systems (SRS) SR830 lock-in amplifier. Differential voltages ($V_{xx}$ and $V_{xy}$) were measured using SRS SR560 voltage preamplifiers with a gain of either 100 or 500 and the signal was band-pass filtered with cutoff frequencies at 10 Hz and 30 Hz and a 6 dB/oct. rolloff. The outputs of the preamplifiers were recorded using SRS SR830 lock-in amplifiers. 2-probe conductance measurements were performed by applying an AC (17.77 Hz) voltage ($\leq$ 200 mV) across adjacent contacts along the same edge. The bias voltage was measured using a SR560 with a gain of 100 and the same filter settings as described above. The current through the device was measured using an Ithaco model 1211 transimpedance amplifier with a gain of $10^{-7}$ A/V.

Four-probe measurements of the 14 nm sample were performed similarly expect that a 100 k$\Omega$ resistor was placed in parallel with the device to minimize Joule heating while measuring the highly resistive state around charge neutrality. Due to the highly insulating state around charge neutrality, measurements of sample 2 where limited to magnetic fields up to 9 T. For all samples, gate voltage, $V_g$, was applied relative to circuit ground and was stepped in 2 mV or 5 mV increments using a Keithley 2400 source meter. The magnetic field was stepped in 200 mT (150 mT for the 14 nm sample) increments. For all data sets, readings taken at the same $V_g$ value were averaged before plotting. Longitudinal ($\sigma_{xx}$) and transverse conductivity ($\sigma_{xy}$) were obtain by tensor inversion as:

$$\sigma_{xx} = \frac{\rho_{xx}}{\rho_{xx}^2 + \rho_{xy}^2} \tag{S8}$$

$$\sigma_{xy} = \frac{-\rho_{xy}}{\rho_{xx}^2 + \rho_{xy}^2} \tag{S9}$$

**X-ray characterization**

X-ray diffraction (XRD) and reflection (XRR) measurements were performed in triple-axis geometry on a Rigaku SmartLab diffractometer equipped with a HyPix3000 detector. The incident optics included a 1 mm physical slit and a Ge (220) 2-bounce monochromator to select Cu $K\alpha$. Figure S2(a) shows x-ray reflectivity measurements for all samples. The sample thickness were determined by fitting the reflectivity data with a model that included all layers of the heterostructures. Figure S2(b) shows 2$\theta$/$\omega$ scans performed after alignment to the GaSb substrate 004 reflection. At room temperature, because of slight difference in growth conditions of the Al$_{0.45}$In$_{0.55}$Sb buffer layer, the measured Cd$_3$As$_2$ 00$\underline{16}$ reflection for the 14 nm and 22 nm samples is shifted to lower angle (concurrently, the Al$_{0.45}$In$_{0.55}$Sb reflection is shifted to a higher angle) than that of the 18 nm, 19 nm, and 20 nm. Figure S3 and S4 show a reciprocal space map (RSMs) taken around the GaSb 224 reflection for 20 nm and 14 nm samples, respectively. Tables SI and SII report the lattice constants, obtained from the RSMs, of the Cd$_3$As$_2$ film and buffer layers for the 20 nm sample and 14 nm sample, respectively. Peak positions were determined by finding the maximum intensity of each reflection after Gaussian filtering with a standard deviation parameter of 1.5. Uncertainties on the lattice constants were estimated by propagating the uncertainties of the peak position in the standard way. Uncertainties of the peak position were taken to equal the step size of the measurement. At room temperature, the 20 nm sample and 14 nm sample are under small compressive strain (-0.54% and -0.75%, respectively) relative to bulk crystals [5].

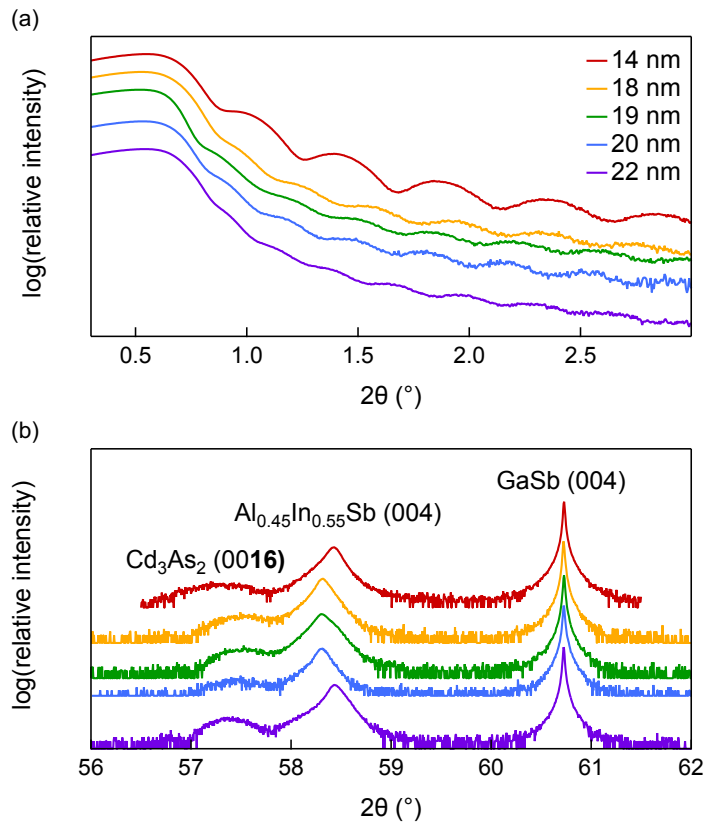

**Fig. S2.** X-ray reflectivity data (a) and out-of-plane 2θ/ω x-ray diffraction patterns (b) of the samples discussed in the manuscript.

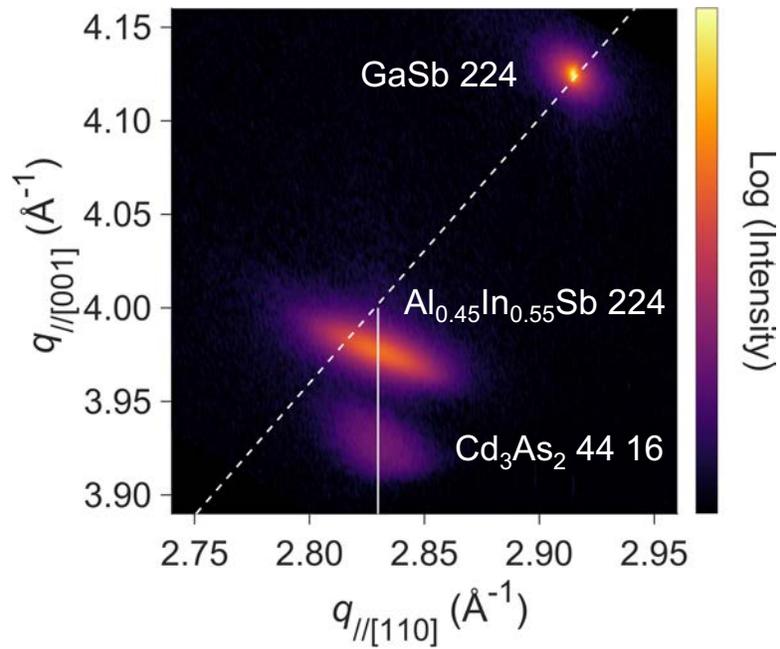

**Fig. S3.** Reciprocal space map taken around the GaSb 224 reflection of 20 nm sample. The dashed white line shows the cubic condition ($a = c$ or $q_{[110]} = \sqrt{2}\, q_{[100]}$) and the solid white line marks the $q_{[110]}$ position of the Al$_{0.45}$In$_{0.55}$Sb buffer layer.

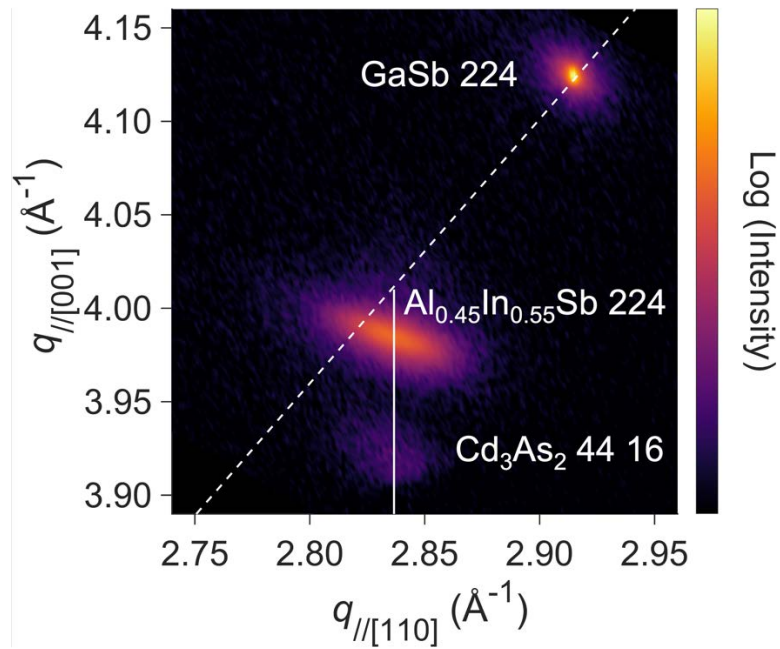

**Fig. S4.** Reciprocal space map taken around the GaSb (224) reflection for the 14 nm sample. The dashed white line shows the cubic condition ($a = c$ or $q_{[110]} = \sqrt{2}\, q_{[100]}$) and the solid white line marks the $q_{[110]}$ position of the Al$_{0.45}$In$_{0.55}$Sb buffer layer.

**Table SI.** Lattice constants of $Cd_3As_2$ and $Al_{0.45}In_{0.55}Sb$ obtained from the RSM in Fig. S3.

| Layer | $a$ (Å) | $c$ (Å) |
|---|---|---|
| $Cd_3As_2$ | 12.565±0.001 | 25.575±0.007 |
| $Al_{0.45}In_{0.55}Sb$ | 6.280±0.001 | 6.319±0.002 |

**Table SII.** Lattice constants of $Cd_3As_2$ and $Al_{0.45}In_{0.55}Sb$ obtained from the RSM in Fig. S4.

| Layer | $a$ (Å) | $c$ (Å) |
|---|---|---|
| $Cd_3As_2$ | 12.538±0.001 | 25.62±0.01 |
| $Al_{0.45}In_{0.55}Sb$ | 6.265±0.001 | 6.309±0.003 |

## Resistance data

Figures S5, S6, S7, S8, and S9 show the longitudinal resistance ($R_{xx}$) and Hall resistances ($R_{xy}$) used to compute longitudinal ($\sigma_{xx}$) and Hall ($\sigma_{xy}$) conductivities for the samples, respectively.

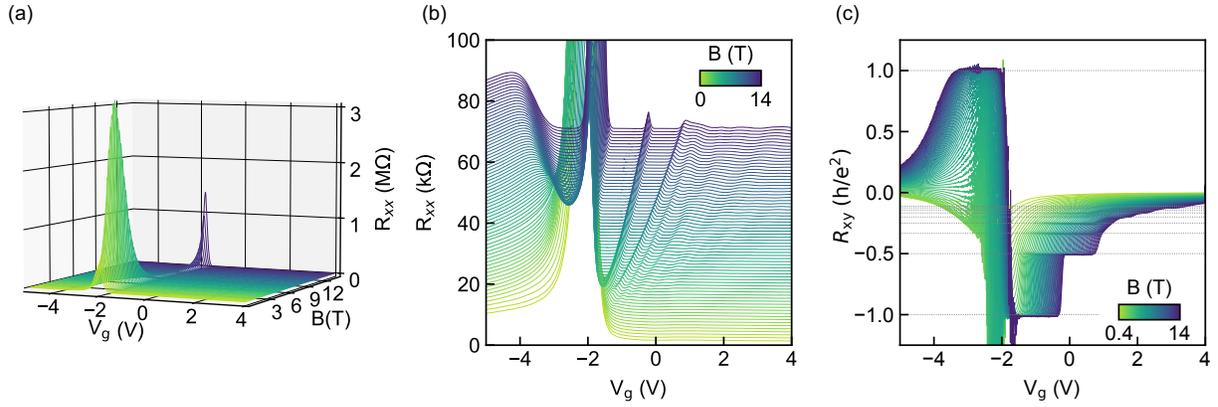

**Fig. S5.** Longitudinal resistance, $R_{xx}$, (a,b) and Hall resistance, $R_{xy}$ (c) of the 20 nm sample. Traces in (b) are offset by 1 k$\Omega$.

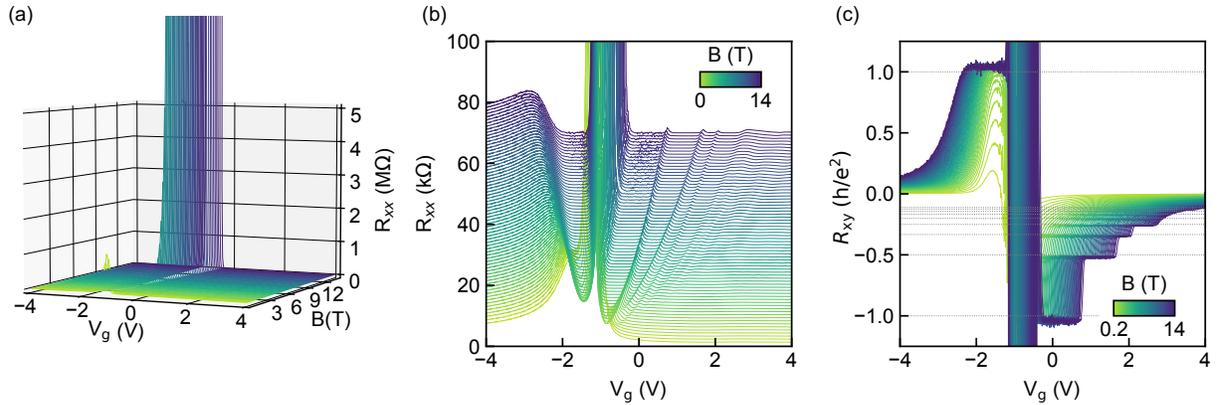

**Fig. S6.** Longitudinal resistance, $R_{xx}$, (a,b) and Hall resistance, $R_{xy}$ (c) of the 18 nm sample. Traces in (b) are offset by 1 k$\Omega$.

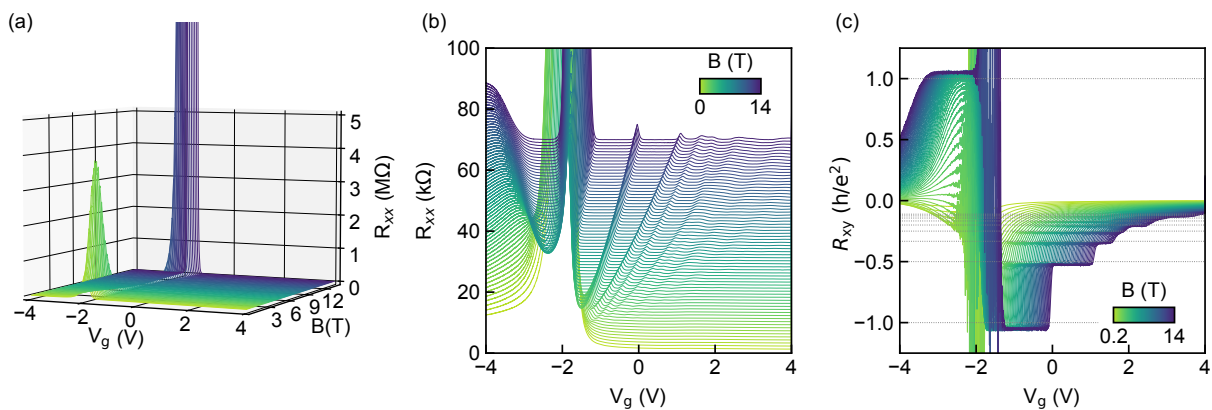

**Fig. S7.** Longitudinal resistance, $R_{xx}$, (a,b) and Hall resistance, $R_{xy}$ (c) of the 19 nm sample. Traces in (b) are offset by 1 kΩ.

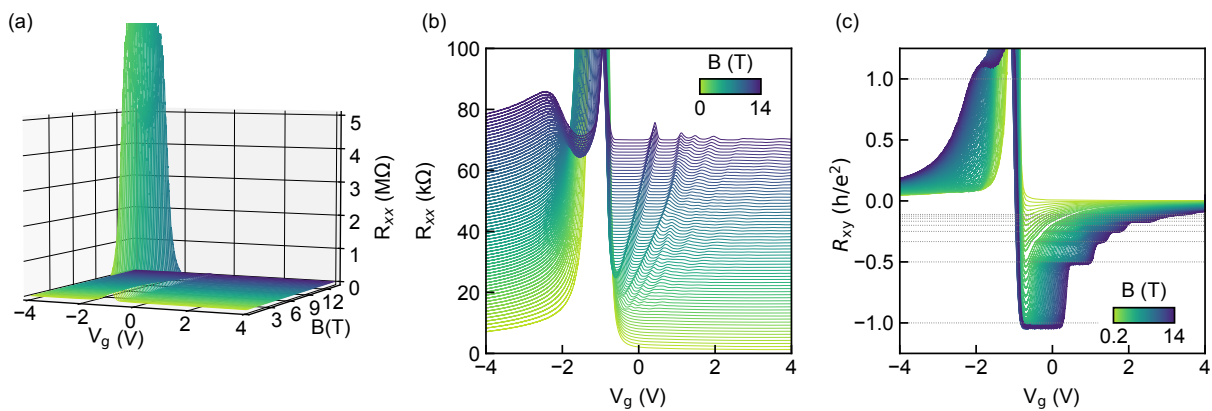

**Fig. S8.** Longitudinal resistance, $R_{xx}$, (a,b) and Hall resistance, $R_{xy}$ (c) of the 22 nm sample. Traces in (b) are offset by 1 kΩ.

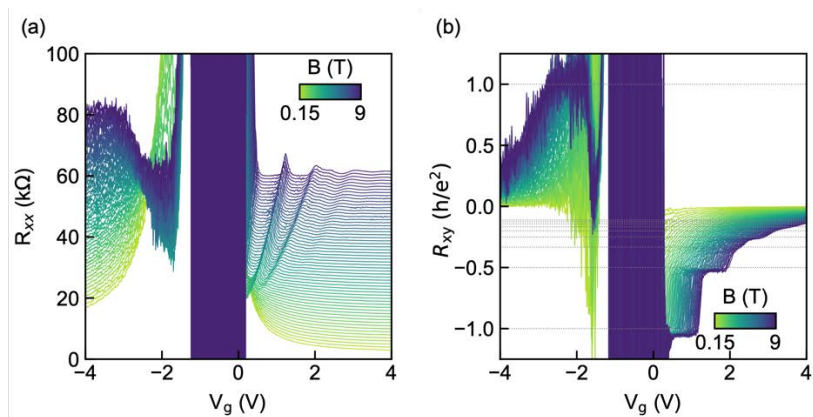

**Fig. S9.** Longitudinal resistance, $R_{xx}$, (a,b) and Hall resistance, $R_{xy}$ (c) of the 14 nm sample.

## Gate voltage dependence of sheet carrier density and Hall mobility

Figure S10 shows the gate voltage, $V_g$, dependence of the sheet carrier density, $n_{2D}$, and the Hall mobility of the 20 nm sample. Capacitance density of the device was determined from the slope of the $n_{2D}$ versus $V_g$. There is a distinct change in slope at $V_g = 1$ V, the approximate voltage where an additional subband appears [see main text as well as Fig. S12(a)], reflecting the influence of the quantum capacitance with the change in density of states. Below $V_g = 1$ V the capacitance density is 62 nF/cm$^2$ and above $V_g = 1$ V the capacitance density is 131 nF/cm$^2$. Assuming a dielectric constant of the 25 nm thick $Al_2O_3$ gate dielectric of 9, the geometrical capacitance is 320 nF/cm$^2$. The discrepancy is due to the influence of the quantum capacitance in series with the gate dielectric.

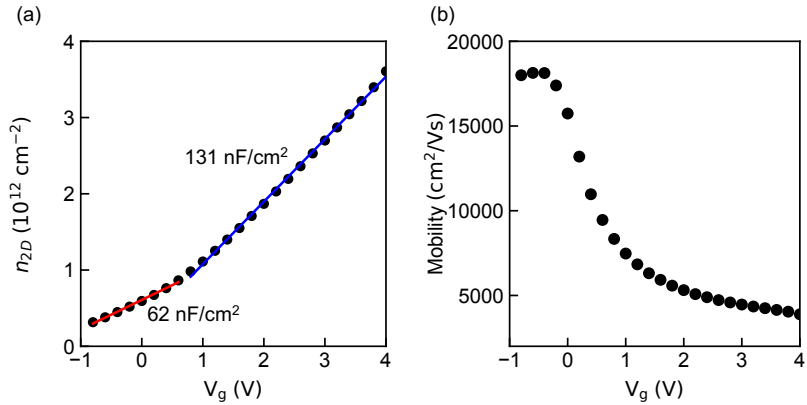

**Fig. S10.** Sheet carrier density, $n_{2D}$, (a) and Hall mobility (b) of the 20 nm sample. The red and blue lines in a are linear fits used to determine the capacitance density.

## Field dependence of $\nu = 0$

Figure S11 shows $\sigma_{xy}$ of sample 1 at 4 T, 9.6 T, and 14 T around $\nu = 0$. For 4 T and 14 T there is a clear plateau at $\sigma_{xy} = 0$, whereas at 9.6 T $\sigma_{xy}$ changes sign smoothly between the $\nu = 1$ and $\nu = -1$ plateaus.

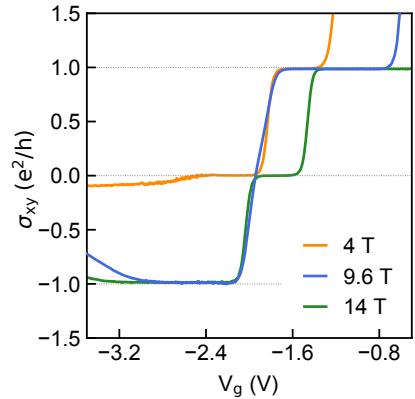

**Fig. S11.** $\sigma_{xy}$ of sample 1 at 4 T, 9.6 T, and 14 T around $\nu = 0$.

## $\sigma_{xx}$ of the 20 nm sample and the 14 nm sample with guides to the eye

Fig. S12 shows $\sigma_{xx}$ for sample 1 (a) and sample 2 (b) with lines overlayed to highlight the Landau levels. The lines serve as guides of the positions of the Landau levels only (they are not fits), as they neglect the

$\sqrt{B}$ dispersion of the $n > 0$ Landau levels. The data show in Fig. S6 is the same as that in Fig. 2(a) and Fig. 4(a) of the main text.

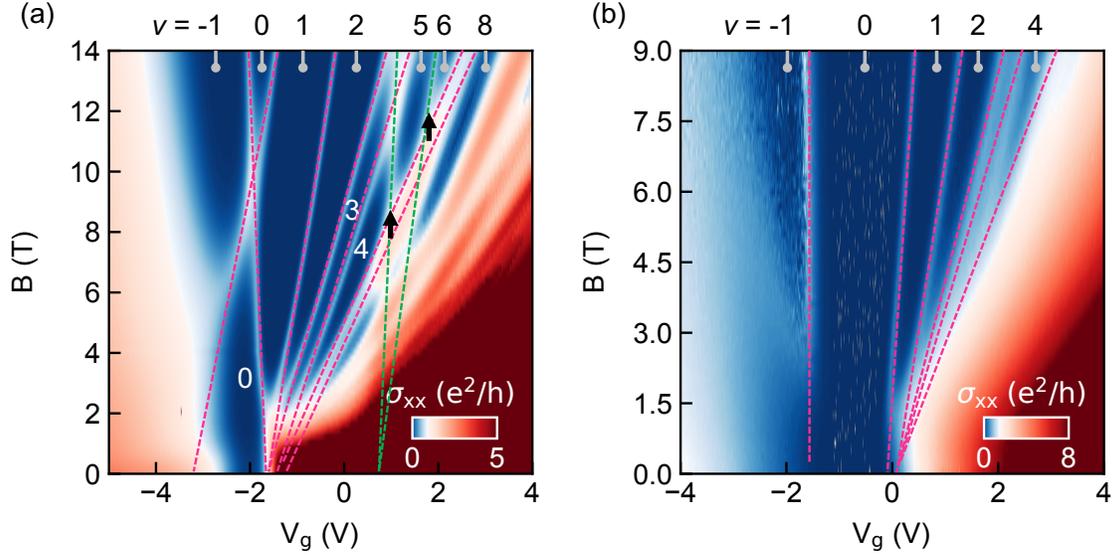

**Fig. S12.** $\sigma_{xx}$ of sample 1 (a) and sample 2 (b). The pink dashed lines are guides to eye for the Landau levels originating from the states nearest to charge neutrality and the green dashed lines in (a) are guides to the eye for the two Landau levels originating from a higher energy subband.

**Temperature dependence of $G$ at $\nu = 0$**

Figure S13 shows temperature dependent two-point conductance of the 20 nm sample. Fitting of the temperature dependence of $G$, shown in Fig. 3 in the main text, was done using the non-linear least squares method as implemented in the SciPy Python package [6]. Uncertainties on the fitted parameters are taken to be equal to the square root of the diagonal elements of the covariance matrix. Tables SII and SIII show the results for the fit parameters.

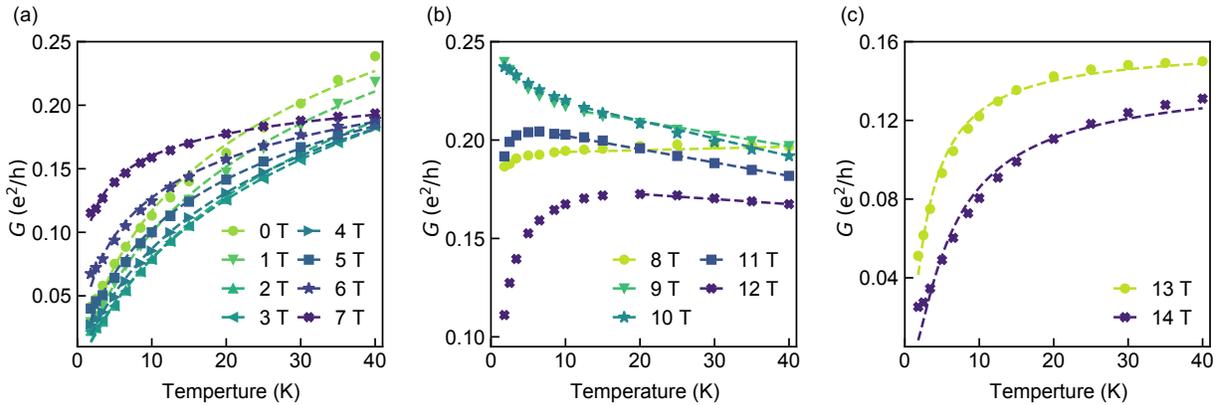

**Fig. S13.** Temperature dependent conductance ($G$) of the 20 sampleat different magnetic fields. The dashed lines in (a) and (c) are fits to the 2D Mott variable range hopping equation and Arrhenius equation, respectively. The dashed lines in (b) are guides to the eye highlighting the range of $T$-linear conductance.

**Table SIII**. Parameters obtained from fitting the minimum 2-point conductance ($G$) at $\nu = 0$ versus temperature to the 2D Mott variable range hopping model, $G(T) = G_0 \exp[-(T_0/T)^{1/3}]$. The experimental data and fits are shown in Fig. S13(a).

| Magnetic Field (T) | $G_0$ (e²/h) | $T_0$ (K) |
|---|---|---|
| 0 | 0.70±0.04 | 58±7 |
| 1 | 0.75±0.03 | 82±6 |
| 2 | 0.81±0.03 | 125±8 |
| 3 | 0.74±0.03 | 112±7 |
| 4 | 0.64±0.02 | 78±5 |
| 5 | 0.50±0.02 | 40±3 |
| 6 | 0.74±0.01 | 11±1 |
| 7 | 0.264±0.003 | 1.2±0.1 |

**Table SIV**. Parameters obtained from fitting the minimum 2-point conductance at $\nu = 0$ versus temperature to the Arrhenius equation, $G(T) = G_0 \exp[-E_a/T]$. The experimental data and fits are shown in Fig. S13(c).

| Magnetic Field (T) | $G_0$ (e²/h) | $E_a$ (K) |
|---|---|---|
| 13 | 0.158±0.002 | 4.8±0.2 |
| 14 | 0.143±0.005 | 10.3±0.8 |

## Magnetic field dependence of $\sigma_{xx}$ between the $n = 0$ Landau levels

Figure S14 shows the conductivity between the $n = 0$ Landau levels versus magnetic field. For $B > 11$ T and $B < 7.8$ T, the $n = 0$ are well separated and there is a clear minimum. Around the crossing of the $n = 0$ Landau levels (11 T $\geq B \geq$ 7.8 T) there is no gap between the $n = 0$ Landau levels and the maximum conductivity is plotted instead.

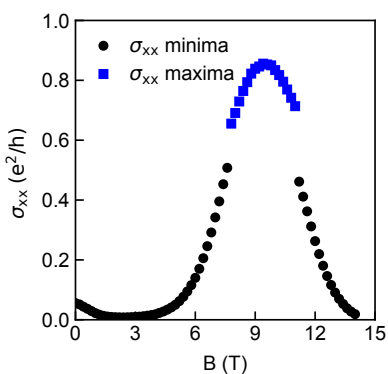

**Fig. S14.** Conductivity of sample 1 between the $n = 0$ Landau levels versus magnetic field.